\documentclass[a4paper,11pt,natbib]{solarphysics}
\usepackage[optionalrh]{spr-sola-addons} 
\usepackage{graphicx}        
\usepackage{url}             

\newcommand{\degr}{^{\circ}}

\newcommand{\aap}{    {\it Astron. Astrophys.}}
\newcommand{\aaps}{   {\it Astron. Astrophys. Suppl.}}

\newcommand{\apj}{    {\it Astrophys. J.}}
\newcommand{\apjl}{   {\it Astrophys. J. Lett.}}
\newcommand{\apjs}{   {\it Astrophys. J. Suppl.}}

\newcommand{\grl}{    {\it Geophys. Res. Lett.}}

\newcommand{\solphys}{{\it Solar Phys.}}

\newcommand{\ssr}{    {\it Space Sci. Rev.}}

\begin{document}

\begin{article}

\begin{opening}

\title{\emph{Solar TErrestrial Relations Observatory}-A and \emph{PRoject for On-Board Autonomy 2} Quadrature Observations of Reflections of three EUV Waves from a Coronal Hole}

\author{I.W.~\surname{Kienreich}$^{1}$\sep
        N.~\surname{Muhr}$^{1,\ 2}$\sep
        A.M.~\surname{Veronig}$^{1}$\sep
        D.~\surname{Berghmans}$^{3}$\sep
        A.~\surname{De~Groof}$^{4}$\sep
        M.~\surname{Temmer}$^{1}$\sep
        B.~\surname{Vr\v{s}nak}$^{2}$\sep
        D.B.~\surname{Seaton}$^{3}$\sep
       }
\runningauthor{Kienreich et al.}
\runningtitle{Reflections of EUV Waves from a Coronal Hole}

   \institute{$^{1}$ Kanzelh\"ohe Observatory/IGAM,Institute of Physics, University of Graz, Universit\"atsplatz 5, A-8010 Graz, Austria
                     email: \url{ines.kienreich@uni-graz.at} email: \url{nicole.muhr@uni-graz.at} email: \url{asv@igam.uni-graz.at} email: \url{manuela.temmer@uni-graz.at}\\
              $^{2}$ Hvar Observatory, Faculty of Geodesy, Ka\v{c}i\'ceva 26, 1000 Zagreb, Croatia
                     email: \url{bvrsnak@gmail.com} \\
              $^{3}$ Royal Observatory of Belgium, Ringlaan 3, B-1180 Brussels, Belgium
                     email: \url{david.berghmans@oma.be} \\
              $^{4}$ ESA, Science and Robotic Exploration Directorate c/o,
                     Royal Observatory of Belgium, Ukkel, Belgium
                     email: \url{anik.degroof@esa.int} \\
                     }

\begin{abstract}
We investigate the interaction of three consecutive large-scale coronal waves with a polar coronal hole, simultaneously observed on-disk by the \emph{Solar TErrestrial Relations Observatory} (STEREO)-A spacecraft and on the limb by the \emph{PRoject for On-Board Autonomy 2} (PROBA2) spacecraft on 27 January 2011. All three extreme-ultraviolet (EUV) waves originate from the same active region, NOAA 11149, positioned at N30E15 in the STEREO-A field-of-view and on the limb in PROBA2. We derive for the three primary EUV waves starting velocities in the range of $\approx310$\,km\,s$^{-1}$ for the weakest up to $\approx500$\,km\,s$^{-1}$ for the strongest event. Each large-scale wave is reflected at the border of the extended coronal hole at the southern polar region. The average velocities of the reflected waves are found to be smaller than the mean velocities of their associated direct waves. However, the kinematical study also reveals that in each case the ending velocity of the primary wave matches the initial velocity of the reflected wave. In all three events, the primary and reflected waves obey the Huygens--Fresnel principle, as the incident angle with $\approx10\degr$ to the normal is of the same magnitude as the angle of reflection. The correlation between the speed and the strength of the primary EUV waves, the homologous appearance of both the primary and the reflected waves, and in particular the EUV wave reflections themselves suggest that the observed EUV transients are indeed nonlinear large--amplitude MHD waves.
\end{abstract}
\keywords{Shock waves, Coronal Mass Ejections}
\end{opening}
\newpage
\section{Introduction}
     \label{S-Introduction}
  Large-scale disturbances propagating through the solar corona were first imaged by the \emph{Extreme-Ultraviolet Imaging Telescope} (EIT) onboard the \emph{Solar and Heliospheric Observatory} (SOHO) about 15 years ago \citep[\emph{e.g.}][]{moses97,thompson98}. These waves, which became known as ''EIT waves'' are now more generally called extreme-ultraviolet or EUV waves. Since their first detection they have been frequently observed in the EUV wavelength range, which has led to a large number of case studies \citep[\emph{e.g.}][]{thompson98,wills99,delanee99,warmuth01,zhukov04,podladchikova05} and several statistical studies \citep[][]{klassen00,biesecker02,thompson09}. Similar large-scale coronal transients were also detected in other wavelengths such as soft X-rays \citep[\emph{e.g.}][]{khan02,hudson03,vrsnak06}, microwaves \citep{white05}, and the metric domain \citep{vrsnak06}.

  Until recently, however, the investigation of large-scale coronal waves has been limited by the low imaging cadence of EIT ($\approx12$ minutes). This resulted in a drastic undersampling particularly of fast events, where at most one single EUV wave front was recorded. The launch of the \emph{Solar Terrestrial Relations Observatory} \citep[STEREO:][]{kaiser08} twin-spacecraft with its \emph{Extreme Ultraviolet Imager} (EUVI) started a new era in high-cadence observations of the solar corona. For more than four years it has supplied observations of EUV waves with high temporal cadence from two different vantage points and over a large field-of-view up to 1.7 solar radii, providing us with new insights into the generation, evolution and 3D structure of large-scale EUV waves \citep[\emph{e.g.}][]{veronig08,patsourakos09,veronig10,temmer11}.

Yet, despite more than 15 years of continuous studies of EUV wave events, the physical nature and generation mechanism of EUV disturbances is still a delicate issue. At present there are two competing theories trying to interpret the observational facts. One group describes these disturbances as fast--mode magnetosonic waves, driven by the CME expansion and/or the explosive flare energy release \citep[\emph{e.g.}][]{warmuth04a,veronig08,patsourakos09}. In pseudo--wave models these disturbances are considered to be a result of the reorganisation of the coronal magnetic field due to the passage of the expanding CME--flank \citep[\emph{e.g.}][]{delanee99,chen02,attrill07,dai10}. \citet[]{delanee00} suggests that a current shell is formed between the rising CME flux tube and the surrounding fields. Joule heating takes place in this envelope, with the thin surface producing a bright front when projected onto the solar disk. The hypothesis of \citet[]{attrill07} is based on the same concept of magnetic--field reconfiguration due to a CME lift--off. However, they proposed a different formation mechanism of the bright fronts. Continuous reconnection between the expanding CME flux tube and favorably orientated ambient magnetic--field lines causes onward--propagating brightenings at the reconnection sites. Hybrid models combining both wave and pseudo-wave interpretations have also been developed \citep[\emph{e.g.}][]{zhukov04,cohen09,liu10}. For further discussion on the observational characteristics and models of EUV waves we refer to the recent reviews of \cite{wills10}, \cite{gallagher10}, \cite{warmuth10}, and \cite{zhukov11}.

Observations of interactions between large-scale EUV waves and active regions or coronal holes have been occasionally reported since the early days of SOHO. \citet{thompson98} were one of the first to report such an interaction, mentioning the halt of EUV waves at the border of coronal holes (CHs). \citet{veronig06} observed a partial penetration of Moreton waves into the coronal hole area with the wave front perpendicular to the CH border. A number of authors have found that EUV waves were stopped or deflected by active regions \citep[\emph{e.g.}][]{thompson99,delanee99,chen05}. Others reported EUV waves moving along the boundaries of active regions and coronal holes \citep[][]{gopalswamy09,kienreich11}. Simulations treating the EUV waves as fast-mode magnetosonic waves supported these findings, as the model MHD waves underwent strong reflections at the borders of high Alfv\'{e}n velocity areas, such as active regions and coronal holes \citep[\emph{e.g.}][]{wang00,ofman02}.

The most intriguing and well studied EUV wave reflection was recorded on 19 May 2007 by the STEREO/EUVI instruments. Several articles report a strong reflection and refraction of an EUV wave at a coronal hole \citep[\emph{e.g.}][]{long08,veronig08}. \citet{gopalswamy09} performed a detailed analysis of this event, identifying reflections that occurred at three different coronal holes. \citet{attrill10}, however, questioned these wave reflections, arguing that the detected features were just artifacts in the running difference images, which were misinterpreted as reflected waves. Instead \citet{attrill10} favored the interpretation that a two-part filament eruption had caused the onset of two EUV waves with the southward--propagating wave experiencing a distinct rotation consistent with the helicity of the associated CME.
It is evident from the limited number of articles dealing with EUV wave interactions with coronal structures that there are many unanswered questions regarding this phenomenon. In particular reports of EUV waves stopping and disappearing at CH borders \citep[\emph{e.g.}][]{thompson98,veronig06} or becoming ''stationary'' fronts \citep[\emph{e.g.}][]{delanee99}, complicates their physical interpretation. Thus it it is still a controversial issue, whether genuine EUV wave reflections have been observed so far, or if they are -- at least according to \citet{attrill10} -- just optical illusions.

In this article we present the first quadrature observations of three large-scale EUV waves that were reflected by a coronal hole. We combine simultaneous STEREO-A on-disk observations, recording both the primary and reflected wave fronts, with the \emph{PRoject for On-Board Autonomy 2} \citep[PROBA2:][]{berghmans06} spacecraft observations of the wave propagation along the limb. This gives us the opportunity to compare the three-dimensional structure of the primary and the reflected wave and to study the change in propagation height caused by the reflection process. Our main focus, however, lies on the kinematical analysis of the on-disk signatures of the reflected waves in relation to the kinematics of their primary counterparts.

\section{Data and Methods} 
      \label{S-data}

The three wave events under study occurred on 27 January 2011. We used data from EUVI, part of the \emph{Sun Earth Connection Coronal and Heliospheric Investigation} \citep[SECCHI:][]{howard08} instrument suite onboard STEREO, in conjunction with simultaneously obtained data from the \emph{Sun Watcher using Active Pixels and Image Processing} \citep[SWAP:][]{degroof08,halain10} instrument on PROBA2. At the time of these events STEREO-A (henceforth ST-A) and SWAP were nearly in quadrature,\emph{ i.\,e.} approximately $90\degr$ apart. Accordingly, while ST-A recorded the EUV waves' on-disk evolution, SWAP observed their propagation along the solar limb.

In our study we use the high-cadence STEREO/EUVI imagery in the 195\,\AA\ passband (cadence of five minutes); in 171\,\AA\ and 284\,\AA\ the cadence was only two hours. SWAP's passband peaks at 174\,\AA, hence its images are comparable to those from the EUVI 171\,\AA\ passband.  During this event, SWAP observed with an image cadence of about 85~seconds. The EUVI 195\,\AA\ filtergrams were reduced using the SECCHI\_prep routines available within SolarSoft. The SWAP data were prepared using the tools provided through the SWAP subtree of Solarsoft. For each wave event, we differentially rotated the images to a specific reference time, namely to 9:15\,UT (event 1), 12:15\,UT (event 2) and 20:30\,UT(event 3) respectively. To emphasize the signature of the transient waves we created running ratio (RR) images, dividing each frame by a frame taken ten minutes earlier, as well as post-event base ratio (BR) images, where we divide each frame by a post-event image taken $\approx20$\,minutes after the last observed wave signature of the event. 
Additionally we applied a median filter over five pixels to each EUVI (2k$\times$2k) image and over three pixels to each SWAP (1k$\times$1k) image to increase the visibility of large-scale structures like EUV waves by smoothing out small-scale fluctuations.
\begin{figure}    
   \centerline{\hspace*{0.01\textwidth}
               \includegraphics[width=0.98\textwidth,clip=]{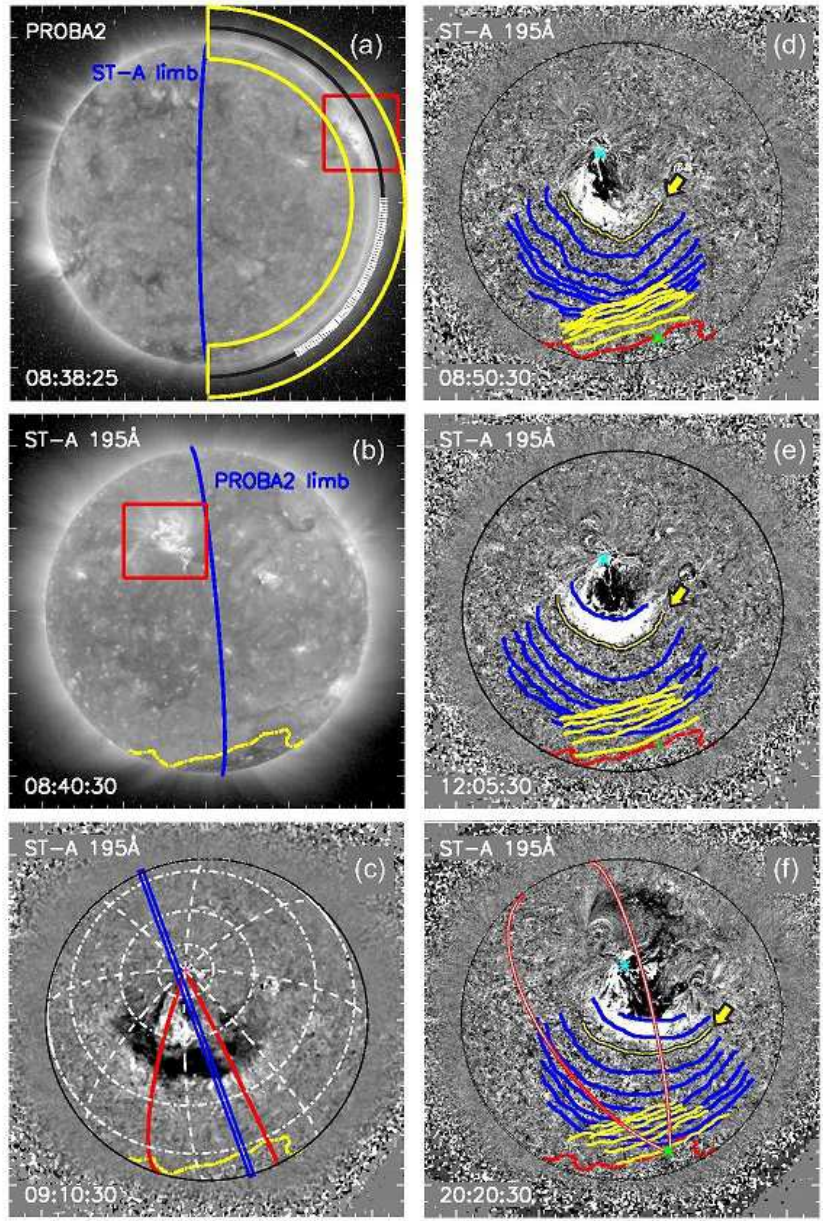}
              }
\caption{(a) PROBA2, (b) ST-A direct image observed on 27 January 2011. The red squares indicate the wave's source region within Active Region NOAA 11149 as seen from both spacecraft. (c)--(f) Median-filtered ten minutes ST-A 195\,\AA\ RR images plus outlines of the southern coronal--hole (dash--dotted line). (c) Red meridians give the delimiting $45\degr$ sector for the calculation of the primary waves' kinematics. (d)--(f) EUV wave events (1)--(3) presenting the primary (blue) and reflected (yellow) wavefronts. The yellow arrow indicates the tracked front associated with the displayed image. (f) Red meridians define the $45\degr$ sector used for the analysis of the reflected wave kinematics. Panels (a) and (c) -- Sectors for stack plots: In (c) the blue rectangle indicates the slice position used for the on-disk stack plots of Figure~\ref{fig6}, and in (a) the yellow semi circle marks the region used for the off-limb stack plots in Figures~\ref{fig7} and~\ref{fig8}.
        }
   \label{fig1}
   \end{figure}
\section{Results} 
      \label{S-results}

\subsection{General Wave Characteristics} 
  \label{S-wavechar}

The three large-scale EUVI waves under study were launched from active region (AR) NOAA 11149 on 27 January 2011 within a period of 12 hours and were all reflected from the same coronal hole near the southern polar region. Each wave launch coincided with a GOES class B/C flare, and the associated coronal mass ejections (CMEs) with plane-of-sky speeds of $455\,\hbox{km\,s}^{-1}$, $413\,\hbox{km\,s}^{-1}$ and $416\,\hbox{km\,s}^{-1}$, respectively, appeared in the LASCO C2 field-of-view (\url{http://spaceweather.gmu.edu/seeds/lasco.php}) at 11:00\,UT, 12:48\,UT, and 20:36\,UT, respectively. ST-A observed the EUV waves on the solar disk with their ejection centers and the sites of the associated flares at the southwestern edge of the AR (N30E15 at onset of the first wave). Note that the positions of the waves' onset centers were almost identical, taking the solar rotation into account. Due to the quadrature configuration of ST-A and SWAP the waves were recorded essentially on the limb by SWAP (Figure~\ref{fig1} (a) and (b)).

Generally EUV waves are best observed in the 195\,\AA\ channel of ST-A, whose response peaks near $T\approx1.4$ MK \citep{wills99}. Nevertheless, they are also clearly visible in the SWAP 174\,\AA\ passband, whose response function has a peak temperature slightly below $T = 1$ MK. Movies 1 and 2 show ST-A and SWAP sequences of direct and RR full--disk images. Since the PROBA2 spacecraft undergoes Earth eclipses during this period of the year, there are data gaps during each event under study. Therefore we do not have a complete set of edge--on observations of the events by SWAP, but in each case the onset of the wave, as well as the entire phase of the wave reflection, was recorded.

All three waves propagate in the same direction and show an angular extent of $\approx120\degr$. As the wave fronts also share similar appearances, these large-scale EUV waves can be considered to be homologous \citep[\emph{cf.}][]{kienreich11,zheng12}. Their similarity is evident in Figure~\ref{fig1}(d)--(f). Each plot displays one representative still frame from the ST-A RR movie of one of the three events, with the locations of the primary wavefronts overplotted in blue. The locations of the reflected wavefronts are shown in yellow. The red dash--dotted contours outline the border of the southern coronal hole (S-CH), where each wave was reflected. We note that the main propagation directions of the primary and reflected waves are each at an angle of $\approx10\degr$ to the meridian perpendicular to the CH border, which implies they obey the laws of reflection. This will be discussed in detail in Section~\ref{S-reflect}.

\begin{figure}[htbp]    
   \centerline{\hspace*{0.01\textwidth}
               \includegraphics[width=0.6\textwidth,clip=]{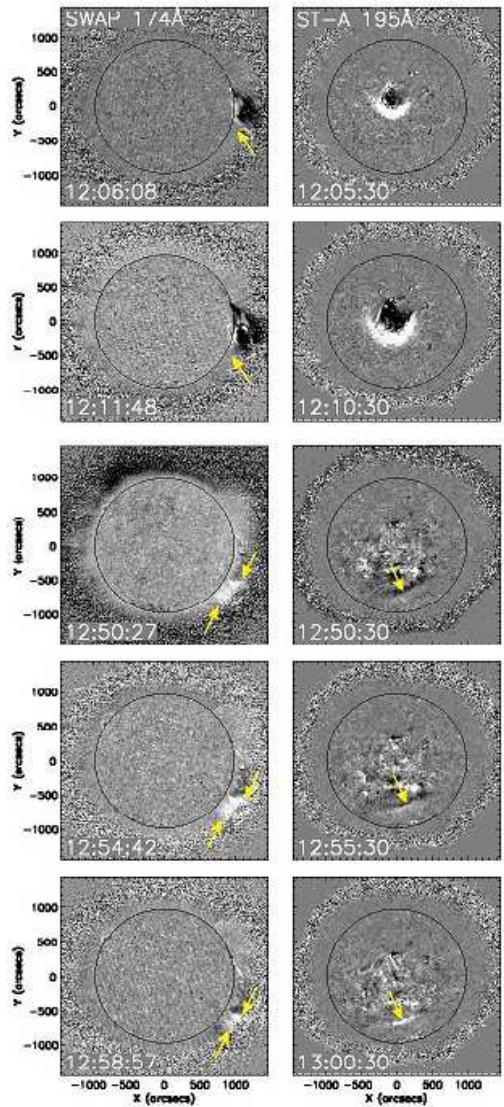}
              }
\caption{Sequence of median-filtered running ratio images of EUV wave event 2 observed at the limb in the SWAP 174\AA\ (left) and on-disk in the EUVI 195\AA\ channel (right). The first two panels show the early evolution of the primary wave, marked by yellow arrows in PROBA2. The last three panels display the evolution of the reflected wave.  Yellow arrows in the ST-A images point to the front of the reflected wave. In contemporaneous SWAP frames the lateral width of the projected wave is indicated by two arrows. See also movie 1 (ST-A) and movie 2 (PROBA2).}\label{fig2}
   \end{figure}

 \subsection{Primary Waves} 
  \label{S-wavekin}

Figure~\ref{fig2} shows the evolution of the second EUV wave event in simultaneous RR snapshots from ST-A and SWAP, extracted from movies 1 and 2. In movie 1 we observed the onset of a rising loop system at $\approx$11:53:00\,UT, before the first wave front appeared at 12:00:00\,UT. Movie 2 gives us the lateral view of the EUV wave event and associated CME. The first CME signature was observed at 11:53:23\,UT exactly on the solar limb at a position of $\approx75\degr$ from the North with its main propagation direction at an angle of $\approx135\degr$ to the North. The first coronal--wave transient appeared in the SWAP field-of-view at 12:00:28\,UT. In the following frames we could discern the wave front propagating close to the limb.

 We tracked the wavefronts manually until 12:40\,UT, thereafter the wave became too weak. The first two panels in each column of Figure~\ref{fig2} illustrate the early phase of the primary wave, starting five minutes after the first wavefront was observed. While the wavefronts are clearly discernible in the 195\,\AA\ ST-A images, they are rather faint in the SWAP 174\,\AA\ recordings (Figure~\ref{fig2}(left); yellow arrows). The SWAP images also show the erupting CME structure, which is highly asymmetric, pointing strongly to the South. The last three images in each column display the evolution of the reflected wave and will be discussed in Section \ref{S-reflect}.

  The determination of the wavefront distances is based on the visual tracking method, applied to the series of ST-A RR images. The wavefronts are defined as the foremost position of the EUV wave signature seen in the RR images using an intensity ratio range of $[0.85,1.15]$. These measurements were counterchecked using BR images with the same intensity ratio range, which delivered essentially the same results. In order to study the wave kinematics, we analyzed each of the three wave events individually. Additionally, we investigated for each event the primary and the reflected wave separately.

  As we were particularly interested in the wave reflection, we concentrated on subsections of the wave fronts propagating towards the southern coronal hole. Hence we considered only those parts of the fronts that lay inside a chosen $45\degr$ sector pointing to the South. This sector is indicated by the red meridians in Figure~\ref{fig1}(c). The intersection point of the two red lines is the ''onset center'' of the wave, which was derived by employing circular fits to the first two tracked wave fronts. This calculation, as well as the determination of the wave front distances from the wave center, is carried out in 3D spherical coordinates; for details we refer to \citet[]{veronig06}.

\begin{figure}[htbp]    
   \centerline{\hspace*{0.01\textwidth}
               \includegraphics[width=0.70\textwidth,clip=]{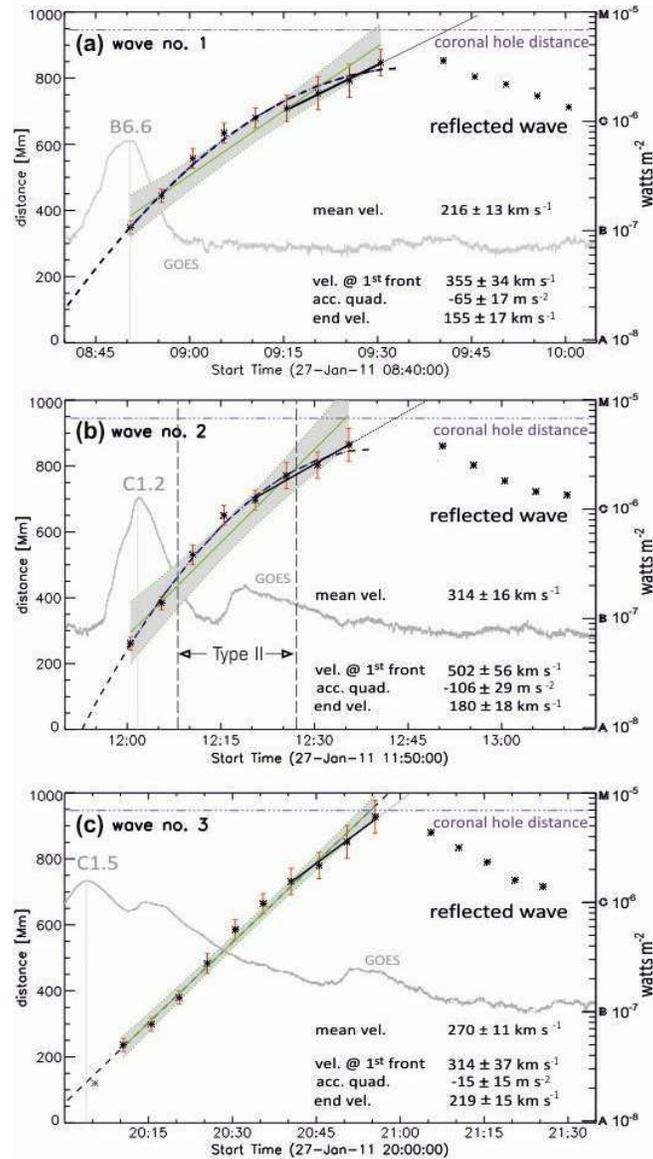}
              }
\caption{Kinematics of primary waves plus error bars. The distances from the wave onset center were derived for a $45\degr$\ sector (\emph{cf.} Figure~ \ref{fig1}(c)). The distances of the reflected wave fronts with respect to the same center and sector are also shown. A linear (green) and a quadratic (blue) least square fit to the primary wave data are overlaid. The grey area indicates the 95\% confidence interval of the linear fit. The dash dotted line in each diagram denotes the average distance of the coronal hole boundary. The velocities derived from the second order polynomial fit are specified for the first observed wave front. The end velocity of each primary wave was obtained by applying a linear fit to the last four data points (black solid line). Grey curves represent the GOES 1 -- 8\AA\ soft X--ray lightcurves of the associated flares.
 }\label{fig3}
   \end{figure}

   Figure~\ref{fig3} shows the time--distance diagrams of all three wave events. The distance values and error bars were statistically derived from five-times tracking of the wave fronts and correspond to the mean values and standard deviations, respectively. The dash--dotted line at $\approx950\,\mathrm{Mm}$ outlines the mean distance of the S-CH border. The coronal--hole border was determined using an automatic algorithm to extract coronal--hole areas \citep[]{rotter12}. In the case of the third event a wavefront was already detected at 20:05\,UT. As the wave signature could not be fully disentangled from the signature of the emerging--loop structure in the RR image, we excluded the corresponding distance value. Taking this uncertainty into account, in all three cases the observation of the first wavefront roughly coincides with the peak time of the associated GOES class B/C flares (see also movie 1) within $\pm$ two minutes. This actually indicates that the associated flare occurred too late to act as the driver of the EUV wave, as it takes some time until the amplitude of the disturbance is large enough to be identified in the RR images \citep[]{veronig08}. On the other hand, the wave--onset times derived from the quadratic fits to the waves' time--distance plots are very well in agreement with the observation times of the first CME signatures seen in movies 1 and 2. We found for EUV wave event 2, for example, an onset time of 11:52:40\,UT, which fits to the times -- namely 11:53:00\,UT (ST-A) and 11:53:23\,UT (PROBA2) -- of the first observed CME signatures. In this respect, it might also be important to note that a type II burst was observed by the Radio Solar Telescope Network observatories San Vito and Palehua, which is closely associated in time with the second, strongest EUV wave event. This provides evidence that a shock--wave formation has taken place due to the steepening of a large-amplitude magnetosonic wave \citep[\emph{e.g.}][]{vrsnak00,white05,zic08}. No type II burst was associated with event 1 and 3.

\begin{figure}[htbp]    
   \centerline{\hspace*{0.01\textwidth}
               \includegraphics[width=0.70\textwidth,clip=]{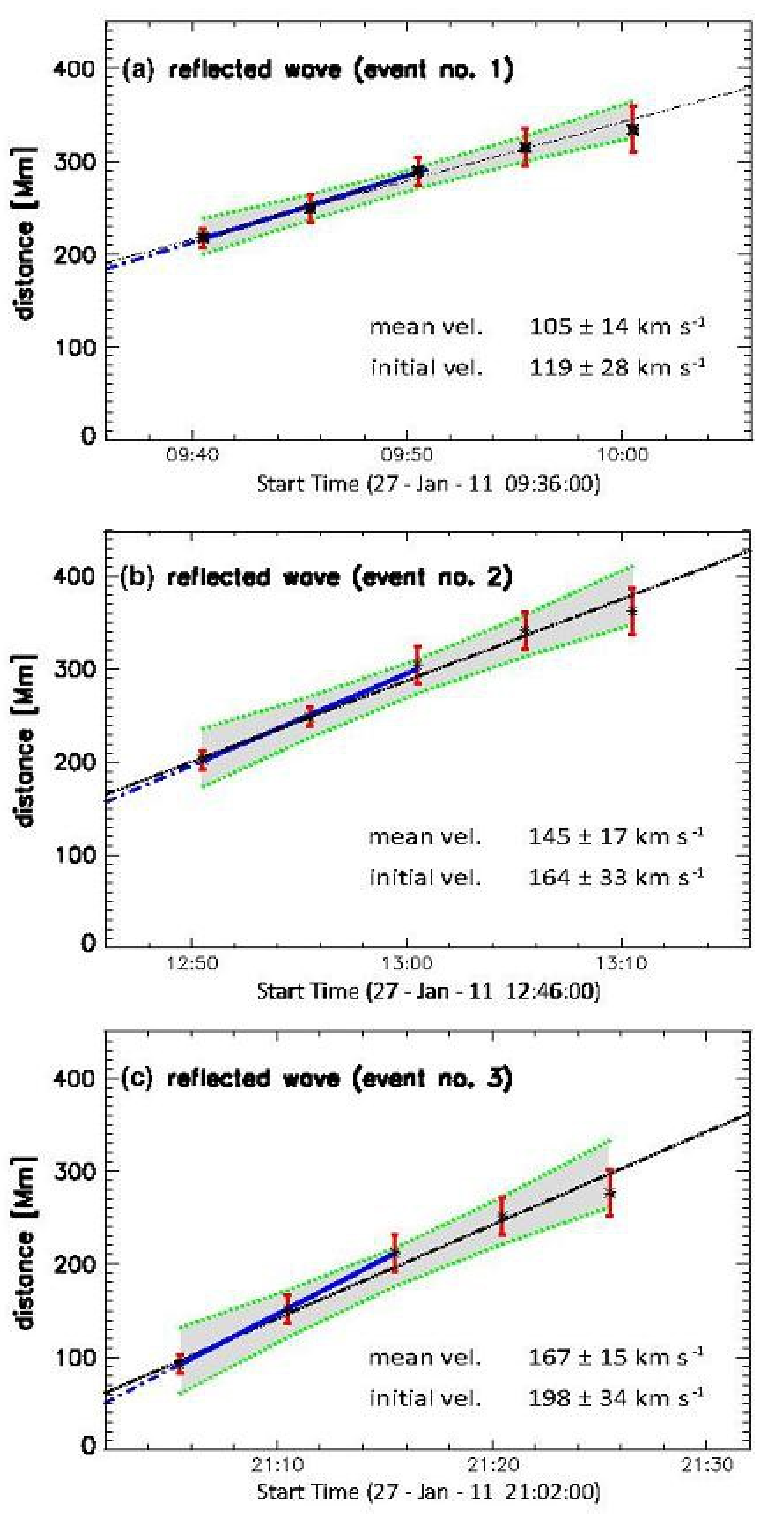}
              }
\caption{Time--distance diagrams of reflected waves together with linear fits and error bars. Grey areas illustrate the 95\% confidence bands for each linear fit. The kinematics of the reflected waves were derived using a reference point at the border of the coronal hole and a $45\degr$ sector (\emph{cf.} Figure~\ref{fig1}(f)) covering the propagation direction of the reflected waves. The blue linear fits to the first three data points deliver the initial velocities of the reflected waves.}\label{fig4}
   \end{figure}

For each event we applied linear as well as least-square quadratic fits to the kinematical curves of the primary wave. Additionally the 95\% confidence interval (Figure~\ref{fig4}; grey areas) was calculated for the linear fit, which gives us a means to decide whether the linear or the quadratic function better represents the wave kinematics. We found the greatest discrepancies in velocity between the two fitting methods for event 2. The linear fit gives a mean wave velocity of $314\pm 16\,\mathrm{km\,s}^{-1}$, while the quadratic fit yields a velocity of $502\pm 56\,\mathrm{km\,s}^{-1}$ at the first observed wave front and a constant deceleration of $106\pm 29\,\mathrm{m\,s}^{-2}$. The derived velocities of event 1 are smaller with $216\pm 13\,\mathrm{km\,s}^{-1}$ resulting from the linear fit and $355\pm 34\,\mathrm{km\,s}^{-1}$ (at the first observed front) from the quadratic fit with a constant deceleration of $65\pm 17\,\mathrm{m\,s}^{-2}$. Event 1 and 2, both show a distinct deceleration, in accordance with previous studies of large-scale EUV waves \citep[\emph{e.g.}][]{veronig08,long08,grechnev11b}. Event 3, however, is consistent with a constant wave propagation with an average velocity of $\approx270\pm 11\,\mathrm{km\,s}^{-1}$. The quadratic fit suggests a small deceleration of $\approx15\pm 15\,\mathrm{m\,s}^{-2}$ and a start velocity of $314\pm 37\,\mathrm{km\,s}^{-1}$. Observations of EUV waves moving at constant speed have already been reported by \citet{ma09}, \citet{kienreich09}.

In order to compare the velocities of the primary waves with the initial velocities of the reflected waves, we also calculated the end velocity of each primary wave by applying a linear fit to the last four data points. The derived end velocity for event 1 is $155\pm 17\,\mathrm{km\,s}^{-1}$, for event 2 $180\pm 18\,\mathrm{km\,s}^{-1}$, and for event 3 $219\pm 15\,\mathrm{km\,s}^{-1}$. We speculate that the reason for the higher ending velocity of event 3 lies in the fact that, as it is the weakest wave, it is barely decelerated. Waves 1 and 2 on the other hand, with larger peak perturbation amplitudes, experienced a strong deceleration. The correlation between these two wave characteristics, the velocity and the perturbation amplitude, is a strong indication for a nonlinear magnetosonic MHD wave behavior of the observed EUV disturbances \citep[\emph{cf.}][]{mann95,vrsnak00}.

\subsection{Reflected Waves} 
  \label{S-reflect}

  In Movie 3 we show a sequence of ST-A RR and direct images, giving a close view of the area in which the reflected wave 2 was observed best. Although faint, the reflected wave can even be detected in the original ST-A images. In Movie 4 the corresponding sequence of BR images is shown. The last three panels of Figure~\ref{fig2} illustrate the propagation of the reflected wave 2. It was first seen at 12:50\,UT and could be followed until 13:10\,UT. The on-disk signatures of the reflected waves, marked by yellow arrows, have an inclination of $\approx20\degr$ with respect to ST-A's North--South axis. Within the same period of time we observed the evolution of an extended bright feature at and above the SWAP solar limb reflecting the EUV wave propagation as seen from ST-A. At 12:50\,UT this bright structure ranged from $\approx123\degr$ to $\approx140\degr$, measured clockwise from SWAP's North. North of it, at $\approx122\degr$, was a narrow brightening -- identified as one leg of a loop connecting to the CME ejection center at $\sim75\degr$-- which did not change its position in time. During the following 20~minutes the width of the wave signature continuously decreased, as its southern boundary moved northwards and ultimately arrived at the CME leg at $\approx122\degr$ at 13:10\,UT.

The derivation of the kinematics for the reflected waves followed the same pattern as previously described for the primary waves. To accurately determine their distances, particularly taking the altered propagation direction into account, we chose a new reference point at the border of the coronal hole and took a different $45 \degr$ sector pointing to the northeast with an inclination of $20\degr$ relative to the solar North--South axis. The new sector of interest is depicted as red--white meridians in Figure~\ref{fig1}(f).

Figure~\ref{fig4} shows the wave kinematics of the three reflected waves. We applied linear fits to all distance points to derive average velocities in the range of $105\pm 14\,\mathrm{km\,s}^{-1}$ (wave 1) up to $167\pm 15\,\mathrm{km\,s}^{-1}$ (wave 3), which were found to be smaller than their primary pendants. All three cases are consistent with a constant propagation within the given errors. To compare the kinematical characteristics of the primary with the reflected waves, we derived the start velocities of the reflected waves by applying a linear fit to the first three data points. In accordance with the end velocities of the primary waves, reflected wave 3 was the fastest with an initial velocity of $198\pm 34\,\mathrm{km\,s}^{-1}$, followed by reflected wave 2 with $164\pm 33\,\mathrm{km\,s}^{-1}$ and reflected wave 1 with $119\pm 28\,\mathrm{km\,s}^{-1}$. The start velocities of the reflected waves are $\approx20\,\mathrm{km\,s}^{-1}$ smaller than the ending velocities of their primary counterparts, which means that the velocities are consistent within the margins of error. These findings strongly support the hypothesis that these EUV events are of the fast-mode magnetosonic wave nature.

We speculate that the small difference between the ending velocities of the primary waves and the initial velocities of the reflected waves can be explained by the fact that the reflected waves move through the flow of the downstream region of the primary waves, which is still directed southward. The downstream flow behind the shock front has a lower speed than the shock itself and extends over a certain distance range behind the shock \citep[]{priest82}. Hence it is conceivable that the downstream region still moves southward while the shock has already been reflected and runs through it. In that case the measured velocity of the reflected wave is a superposition of its true velocity and the oppositely directed flow speed of the downstream region.

\begin{figure}
\centerline{\hspace*{0.01\textwidth}
  \includegraphics[width=0.98\textwidth]{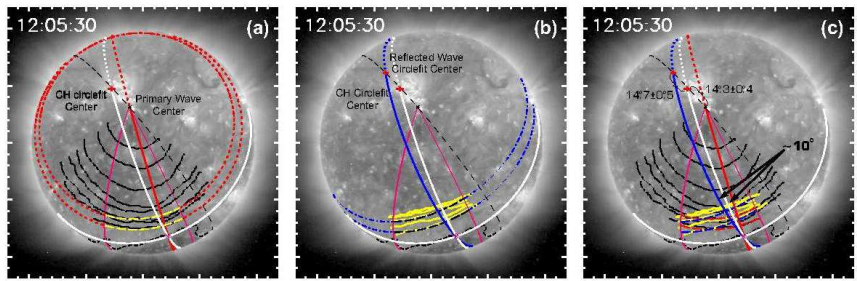}
  }
  \caption{Sketch of the EUV wave event 2 at different times superposed on a STEREO/EUVI 193\,\AA\ direct image overlaid by the coronal--hole border (dashed black line) and delimiting $45\degr$ sector for the calculation of the primary waves' kinematics (pink meridians). The circular fit to the coronal--hole limb and its related perpendicular great circle are displayed as white curves. (a) Circle fits (dashed curves) were applied to the primary wavefronts (black). The perpendicular great circle representing the main propagation direction is shown in red.  (b) Reflected wavefronts (yellow) plus circle fits (dash--dotted lines) and associated main propagation direction (blue). (c) The incident angle with respect to the CH normal is with $\approx10\degr$ of the same size as the angle of reflection.}\label{fig5}
\end{figure}

Figure~\ref{fig5} gives evidence that the observed propagating disturbances obey the Huygens--Fresnel Principle. For this purpose, we compared the angle of incidence of the primary waves and the angle of reflection of the secondary waves with respect to the coronal--hole border. All measurements and calculations were carried out in 3D spherical coordinates. Firstly, we applied a circular fit to the coronal--hole border. The resulting circle fit and the great circle perpendicular to it are displayed as white curves in Figure~\ref{fig5}(a)--(c).  In the same manner circular fits were applied to the wave fronts of the primary wave, \emph{cf.} Figure~\ref{fig5}(a), and to the secondary wave fronts, \emph{cf.} Figure~\ref{fig5}(b). Figure~\ref{fig5}(c) provides the final result for wave event 2. The derived meridians perpendicular to the wave fronts represent the main propagation direction of the primary (red) and reflected wave (blue), respectively. The angular distance between the primary wave center and the CH circle fit center ($14.\!\!\degr3\pm0.\!\!\degr4$) matches the angle between the CH circle fit center and the reflected wave center ($14.\!\!\degr7\pm0.\!\!\degr5$). The angle between the two wave propagation directions measured at the CH vertex (Figure~\ref{fig5}; pink star) is $\approx20\degr$. The angle of incidence and the angle of reflection with respect to the coronal--hole border normal are of the same size with $\approx10\degr$. This is a clear indication that the observed features indeed obey the laws of wave reflection.

We note that the choice of the intersection point at the coronal hole border together with the consequential perpendicular meridian are arbitrary and could be exchanged. Here we used the reference point for the calculation of the reflected wave kinematics. However, each point on the CH border is valid and ultimately leads to the same result. This lies in the fact that, as the angular distance to the CH border is fixed, a change of the CH vertex position certainly results in a change of the total angle between the two propagation directions, but the incident angle still remains of the same size as the reflection angle.

 \begin{figure}[htbp]    
   \centerline{\hspace*{0.01\textwidth}
               \includegraphics[width=0.7\textwidth,clip=]{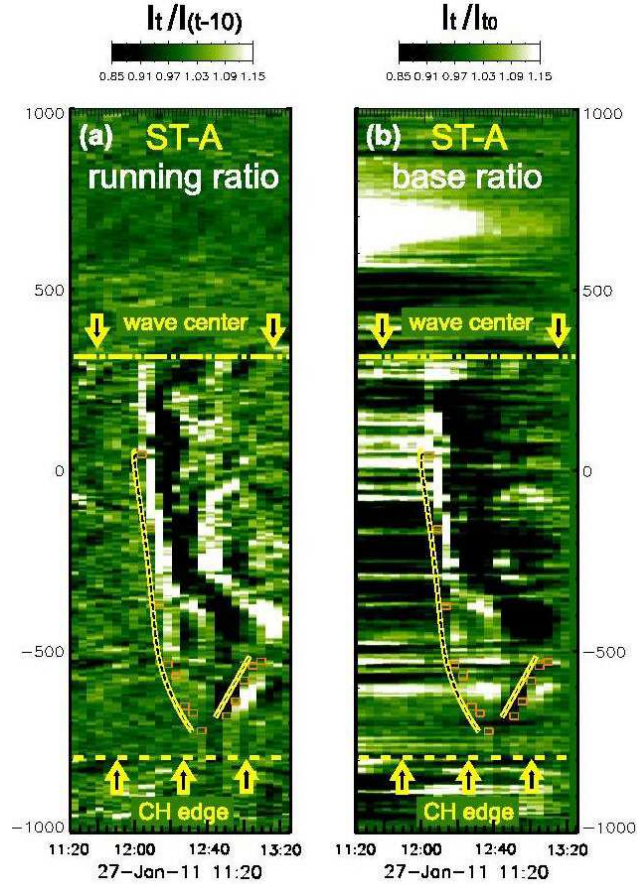}
              }
\caption{(a) On-disk stack-plot of ST-A 195\AA\ RR images.The cut-out region used for the stack-plot is marked as blue rectangle in Figure~\ref{fig1}(c). The wave propagation is outlined by the yellow curves, the primary wave is traced by a ''parabolic'' curve and the reflected wave by a straight line. The dashed line represents the border of the coronal hole, and the dash dotted line the position of the wave center. (b) The reflected wave is also evident in the stack-plot of ST-A 195\AA\ post--event BR cut-outs (base-time: 13:25\,UT). The orange boxes indicate the tracked wave front position at each time index for the given slice.}\label{fig6}
   \end{figure}

\subsection{Stack Plots and Lateral View of the Events} 
  \label{S-lateral}

 In order to discern the reflected wave more clearly we created stack plots from slices cut out from ST-A RR and BR images. Each slice has a width of eight pixels and its vertical axis spans from $-1000''$ to $+1000''$. The position and direction ($20\degr$ to NS) of such rectangular slice is exhibited as blue rectangle in Figure~\ref{fig1}(c). In the resulting stack plot for event 2 (Figure~\ref{fig6}(a)) slices of RR images taken at subsequent time-indices are stacked together to reveal the evolution of wave 2 along the slit. The primary wave appears as a bright front moving downwards between 12:00\,UT and 12:35\,UT, and shows a distinct deceleration. It is traced by the yellow ''parabolic'' curve. The reflected wave is a bright front running upwards, outlined by a straight line. It is visible between 12:50\,UT and 13:10\,UT. To determine whether the observed features, which we assume to be reflected waves, are not just artifacts in the RR images, where a previously dark feature shows itself as bright feature in the next frame and vice versa, we also generated stack plots from post--event BR images. The base image was taken $\approx20$ minutes after the last observed reflected wave front. At that time the background corona had returned to its pre--event quiet Sun state. As it is evident from Figure~\ref{fig6}(b), the reflected wave is recognizable in the BR stack plot, hence the feature is a real physical observable.

 \begin{figure}[htbp]    
   \centerline{\hspace*{0.01\textwidth}
               \includegraphics[width=0.95\textwidth,clip=]{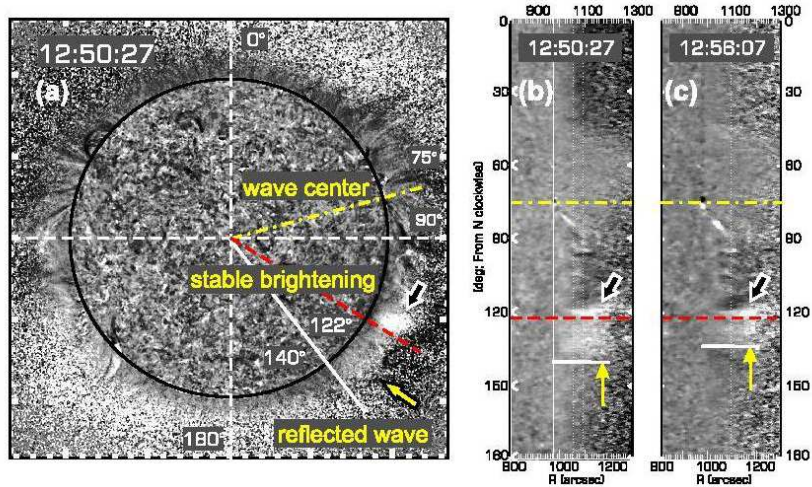}
              }
\caption{(a) SWAP BR image revealing a lateral view of the first observed reflected wave front of event 2 with position angles of the wave center (yellow dash--dotted line), stationary brightening (red dashed line) and southern limb of the EUV wave (white line).(b) and (c) Polar display of two PROBA2 RR-images taken along the solar limb at $R= [800''$ to $1300''$]; $\theta= [0\degr$(North) to $180\degr$(South)] (\emph{cf.} Figure~\ref{fig1}(a) yellow semi-circles). The same lines as used in panel (a) denote the positions of the wave center and stationary brightening (black-white arrows). A comparison of panel (b) and (c) reveals the propagation of the waves' southern limb, indicated by the yellow arrow, towards north. From such polar plot, slices are cut out at different heights covering the full $\theta$--range. The resulting stack-plots for event 2 are shown in Figure~\ref{fig8}.}\label{fig7}
   \end{figure}

In order to compare the ST-A on-disk stack--plots with stack--plots generated from PROBA2 limb images we cut out a semi-circular ring from each PROBA2 BR image. This area is limited to a radial range $R = 800''$ to $1300''$ and an angular range $\theta = 0\degr$ (North) to $180\degr$ (South) with $\theta$ increasing clockwise. In Figure~\ref{fig1}(a) this area is marked by yellow semi-circles. All features lying in this area, which are relevant for wave event 2, are tagged in Figure~\ref{fig7}(a). The wave onset center is at $75\degr$ from the North, the stable brightening stays at a position of $\approx122\degr$, while the southern edge of the EUV wave is at $\approx140\degr$ at the observation time of the displayed BR image (12:50\,UT). The position of these structures are also marked in Figures~\ref{fig7}(b) and \ref{fig7}(c), which are two polar plot representations of the cut-out region of Figure~\ref{fig1}(a). Here, we transformed the cut--out region, originally given in Cartesian $(x,y)$--coordinates, into polar ($R, \theta$)--coordinates; in the resulting plots, $R$ (arcsec) is defined along the horizontal and $\theta$ (degree) along the vertical axis.

\begin{figure}[htbp]    
   \centerline{\hspace*{0.01\textwidth}
               \includegraphics[width=1.5\textwidth,angle=270,clip=]{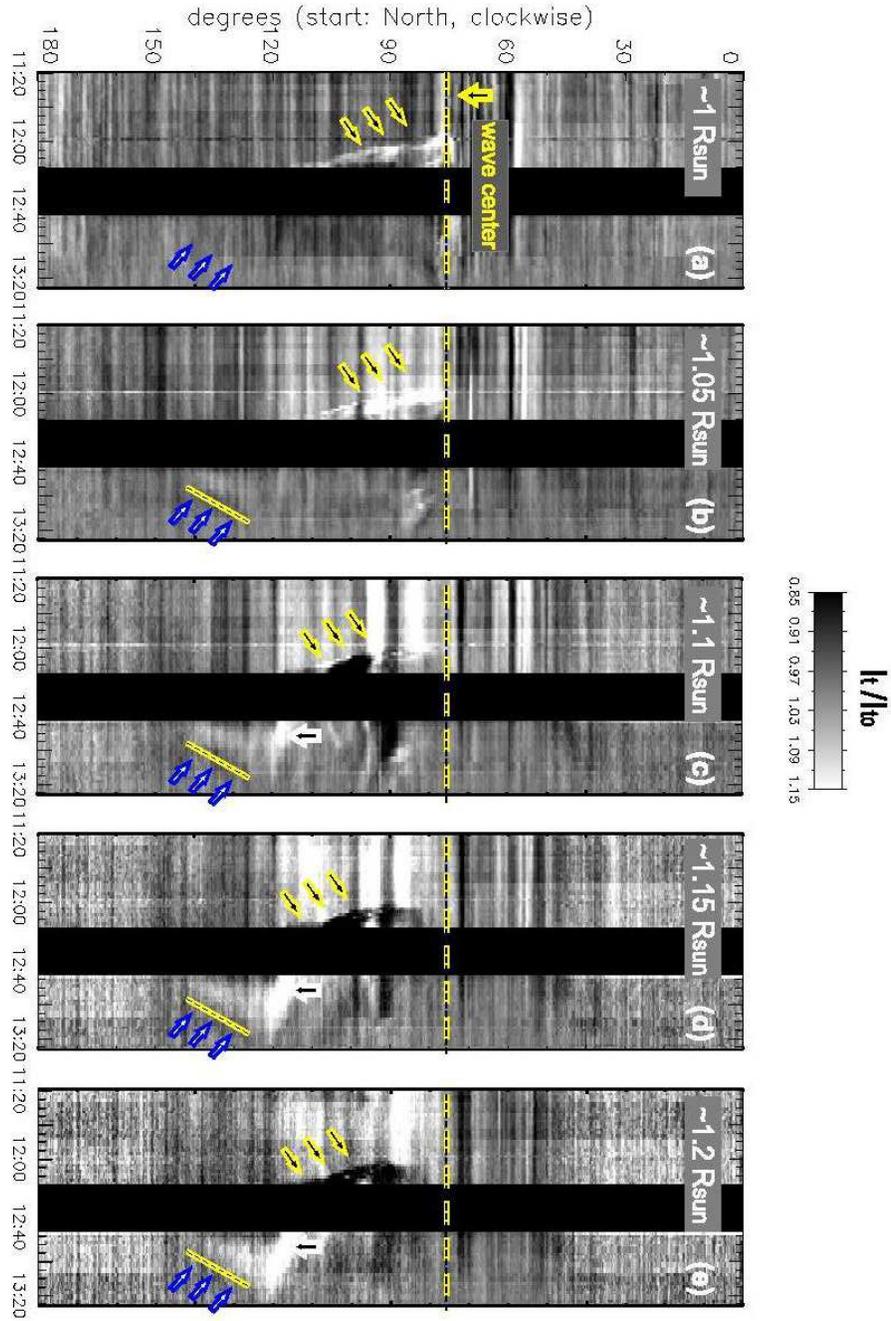}
              }
\caption{PROBA2 BR stack plots for different heights above the solar limb. The signature of the primary wave (yellow arrows) is evident at $R\approx1\,R_{\odot}$ (a) and $R\approx1.05 R_{\odot}$ (b). It is barely visible at larger heights, as shown in (c) to (e). The data gap in all five plots between 12:15\,UT and 12:40\,UT is due to a PROBA2 eclipse. The reflected waves (blue--white arrows) show the strongest signature in a height range of $H\approx0.15$ to $0.2\,R_{\odot}$, (d) and (e), while they are invisible at the solar limb. Blue--white arrows in panel (a) indicate the wave position at $H\approx0.1 R_\odot$.}\label{fig8}
   \end{figure}

Finally, slices at different coronal heights are cut out and stacked together. In Figure~\ref{fig8} we show five stack--plots extracted from slices centered at $0'' (\mathrm{limb})$, $50'' (\approx\!0.05\,R_\odot), 100'' (\approx\!0.1\,R_\odot), 150'' (\approx\!0.15\,R_\odot) \mathrm{\ and\  200'' (\approx\!0.2\,R_\odot)}$ above the limb. It is important to emphasize that the plots in Figure~\ref{fig8} are also created from post--event BR images, thus avoiding any potential introduction of artifacts. In each of the stack--plots the onset of the EUV wave event is evident. However, while we clearly see the signature of the primary wave at lower heights, appearing as bright feature moving to larger $\theta$--values (Figure~\ref{fig8}(a,b) yellow arrows), at larger heights we only recognize a southward propagating dark feature, which we identify as coronal dimming and as such as CME signature (Figure~\ref{fig8}(d,e) yellow arrows). A comparison of the different stack plots reveals an increasing time delay with height in the observation of the wave signature and also of the coronal dimming.  Moreover, an increase in width of the CME signature with height is evident, which is a sign for the expected upward and lateral expansion of a CME. We note that between 12:15\,UT and 12:40\,UT we do not have records from the CME and wave commencement, as PROBA2 experienced an eclipse during this time, but SWAP observed the entire propagation of the reflected wave (Figure~\ref{fig8} blue--white arrows). During the final phase of the observation no further dimmings were observed.

Our detailed analysis of the stack--plots at different heights provides another important result. The stack--plots give evidence that the strongest signatures of the primary wave and reflected wave move at different heights. While the primary wave is brightest at the limb up to $R \approx 1.05\,R_\odot$, it is merely recognizable at $1.1\,R_\odot$ and imperceptible at $R\geq 1.15\,R_\odot$. The signature of the reflected wave however is strongest in a radial distance range of $R = 1.1$ to $1.2\,R_\odot$, the strength of the wave signal decreases to larger as well as to smaller heights. At $1.05\,R_\odot$ the wave signature is very faint and at the limb there is no evidence of the reflected wave at all. These findings are one striking piece of evidence that suggests that the waves were not simply reflected in a two-dimensional plane at a fixed height, but that they were also reflected toward larger coronal heights.

We compared the evolution of the bright feature in the ST-A on-disk slice (Figure~\ref{fig6}) with the evolution of the brightening in the SWAP limb slices (Figure~\ref{fig8}). Our study suggests that the bright feature recorded in the ST-A slice, indicating an upward movement at constant speed, coincides with the bright structure, proceeding northward to smaller $\theta$--values, in the SWAP limb slices at heights of around $1.15$ to $1.2\,R_\odot$.

In order to check the on-disk velocity we estimated the speed of the reflected wave from the stack plot at $R = 1.2\,R_\odot$. The wave covers an angular distance of $\approx17\degr$ in a time interval of $\approx20$\,minutes, which gives a velocity of $v = (\frac{17\cdot\pi}{180}\cdot 1.2\cdot R_\odot)/1200 \approx 206$\,km\,s$^{-1}$, which is by a factor of $\approx1.2$ higher than the measured on-disk velocity. This brings up another hypothesis as to why we derived such low velocities in our on-disk kinematical analysis of the reflected waves. Using ST-A on-disk measurements for our kinematical study, we only observe the wave fronts projected onto the solar surface without any height information. Consequently we derive velocities with an inherent error due to this projection effect \citep[\emph{cf.}][]{kienreich09}. The primary waves are propagating at lower heights of $0.0$ up to $0.05$ solar radii, thus the error is marginal, while the reflected waves propagate at a height of $\approx 0.2 R_\odot$, leading to a larger error. If we compensate for this height difference by multiplying the on-disk velocity by a factor of 1.2, we obtain for the reflected waves start velocities in the range of $\approx 140$\,km\,s$^{-1}$ (wave 1) up to $\approx 220$\,km\,s$^{-1}$ (wave 3), which are consistent with the ending velocities of the primary waves without any offset in speed. The height--corrected velocity of wave 2 yields $\approx 200$\,km\,s$^{-1}$, which matches the velocity derived from the stack plot at $R = 1.2\,R_\odot$.

Moreover, the velocities are consistent with a magnetosonic speed of $160$\,km\,s$^{-1}$ to $230$\,km\,s$^{-1}$ for a propagation height of $\approx0.2\,R_\odot$. The values for the magnetosonic speed were derived by applying three-- to five--fold Saito coronal--density models for a temperature of 1\,MK under the assumption of a magnetic--field strength of $B\approx0.5$\,G to $\approx1.1$\,G at the given height. Here we used the measurements of the quiet Sun coronal magnetic--field strength by \citet[]{lin04} and \citet[]{liu08a}.

\section{Discussion and Conclusion} 
      \label{S-Conclusion}

We have presented observations of the reflection of three homologous coronal EUV waves from the border of a coronal hole observed simultaneously in rapid cadence by ST-A and PROBA2 in quadrature configuration within a period of 12 hours. The on-disk view of ST-A revealed that all three primary large-scale EUV waves were launched from the same onset center and propagated into the same direction. PROBA2 provided an edge--on view of the events, showing the southward directed movement of the primary waves along the limb in a height range between $0$ (limb) and $\approx$\,$\mathrm{0.05}\,R_{\odot}$ above the limb.

 The most important result of our study is the reflection of all three waves from the boundary of the same coronal hole close to the south pole. In each of the three EUV wave events the incident angle was of the same size as the reflection angle, no matter which vertex on the coronal hole border was chosen. Furthermore we found that the initial velocity of the reflected waves are consistent with the end velocities of the incident waves. These two key results confirm that the observed EUV disturbances obey the Huygens--Fresnel Principle, and hence are indeed fast-mode magnetosonic waves. The additional lateral view from PROBA2 revealed a northward directed motion of the reflected waves at a height of $\approx$\,$0.15\,\mathrm{to}\,0.2\,R_{\odot}$ fitting to the on-disk results.

 The reflection of the EUV brightenings at the coronal--hole border is consistent with the characteristics of fast-mode magnetosonic waves, but cannot be expected from CMEs. Theoretically the observed incident waves could be explained as CME signatures. However, at the border of the coronal hole the CME front would not propagate backwards but would be deflected upwards, as the CME would still continue to expand. Apparently, we would then observe a completely different geometry of the EUV brightenings in the final phase of each event. Hence, the observational facts do not support the pseudo-wave models, which are based on a magnetic--field line reconfiguration at the CME front

 Further pieces of evidence of the fast-mode magnetosonic wave interpretation are provided by the following findings:\\
  (i) The primary large-scale EUV waves, launched from the same onset center, propagated into the same quiet--Sun area and displayed a similar appearance and angular extent. This cannot be expected for CME related features, as in this case after one event the background magnetic--field configuration is permanently changed. Their behavior and appearance strongly resembles the homologous events reported by \citet{kienreich11}.\\
  (ii) Primary waves 1 and 2 with initial velocities $v$ of $\approx$\,$350$\,km\,s$^{-1}$ and $\; $  $\approx$\,$500$\,km\,s$^{-1}$, respectively, showed a distinct deceleration, while wave 3 propagated at a constant speed of $\approx$\,$270$\,km\,s$^{-1}$.  The ending velocities are in a range of $\approx$\,$155$\,km\,s$^{-1}$ (wave 1) up to $\approx$\,$220$\,km\,s$^{-1}$ (wave 3). We note that wave 2 is the strongest wave with the largest peak perturbation amplitude $A$, i.\,e. intensity ratio [$I/I_0$], while wave 3 is the weakest wave. This result, [$v \propto A$], is also in favor of the nonlinear large--amplitude MHD wave interpretation \citep[\emph{e.g.}][]{kienreich11}. Moreover the deceleration is also proportional to the wave velocity, [$a \propto v$], which is another indication for the wave interpretation.\\
 (iii) The fastest and strongest primary large-scale wave event 2 was accompanied by a metric type II burst indicative of a coronal shock wave.\\
  (iv) As the coronal--hole boundary did not change noticeably within the 12--hour period, we expect the reflected waves to be equally homologous like the primary waves. In the ST-A images it was indeed obvious that all three reflected waves had the same shape and angular extent and propagated into the same direction, $\approx$\,$20\degr$ inclined to the direction of the primary waves.\\

  As the EUV waves were accompanied by CMEs and GOES B/C flares, we also looked into the timing of these features. In each event we found the flare peak to occur approximately at the time of the appearance of the first wave front. The timing is an argument against a blast--wave scenario, in which the wave is initiated by the explosive flare--energy release, as there should be a delay between the flare peak and the manifestation of the first EUV wave in the ratio images. The timing and direction of the erupting CME however indicates a close association between the large--scale coronal wave and the CME. We conclude that the observed EUV disturbances are fast mode magnetosonic waves initiated by the CME expanding flanks. It is evident from the SWAP plane-of-sky observations (\emph{cf.} movie 2) that the standoff distance between the wave front and the CME increases with time. This is expected from a large-scale coronal wave, which is driven only for a short time and then propagates freely.

\begin{acks}
 I.W.K., N.M., and A.M.V. acknowledge the Austrian Science Fund (FWF): P20867-N16 and P24092-N16. The MOEL F\"orderungsprogramm is acknowledged by N.M. The European Community's Seventh Framework Programme (FP7/ 2007-2013) under grant agreement no. 218816 (SOTERIA) is acknowledged (B.V., M.T.). We thank the PROBA2 team and STEREO/SECCHI teams for their open data policy. SWAP is a project of the Centre Spatial de Li\`{e}ge and the Royal Observatory of Belgium funded by the Belgian Federal Science Policy Office (BELSPO). I.W.K. thanks the PROBA2 team members for their support during her stay at the Royal Observatory of Belgium as PROBA2 guest investigator.
\end{acks}

\end{article}
\bibliographystyle{apj}

\end{document}